\documentclass[12pt,]{article}
\usepackage{authblk}
\usepackage{fullpage}
\usepackage{amssymb,amsmath}
\usepackage[utf8x]{inputenc}
\usepackage[T1]{fontenc}
\usepackage{siunitx}
\usepackage[version=3]{mhchem}
\usepackage{url}

\usepackage[sort&compress,square,comma,numbers]{natbib}
\usepackage{setspace}
\usepackage{graphicx}
\graphicspath{{figures/}}
\makeatletter
\def\ScaleWidthIfNeeded{%
 \ifdim\Gin@nat@width>\linewidth
    \linewidth
  \else
    \Gin@nat@width
  \fi
}
\def\ScaleHeightIfNeeded{%
  \ifdim\Gin@nat@height>0.9\textheight
    0.9\textheight
  \else
    \Gin@nat@width
  \fi
}
\makeatother
\setkeys{Gin}{width=\ScaleWidthIfNeeded,height=\ScaleHeightIfNeeded,keepaspectratio}%

\title{Experimental demonstration of a beam shaping non-imaging metasurface}
\author[1]{Kirstine E. S. Poulsen}
\author[1]{Xavier Zambrana-Puyalto}
\author[2]{Rafael de la Fuente Herrezuelo}
\author[2]{Villads Egede Johansen}
\author[1,*]{S\o ren Raza}
\affil[1]{Department of Physics, Technical University of Denmark, Fysikvej, DK-2800 Kongens Lyngby, Denmark}
\affil[1]{NIL Technology ApS, Haldor Topsøes Allé 1, DK-2800 Kongens Lyngby, Denmark}
\affil[*]{sraz@dtu.dk}

\date{}

\begin{document}

\maketitle

\newpage
\section*{Abstract}
Light emitting diodes have superior performance over most other light sources, but the need for secondary optics to shape their illumination for specific applications yield bulky lighting products. Here, we present an approach to shaping light from incoherent sources, such as light emitting diodes, using non-imaging metasurfaces. We present a theoretical framework and a numerical tool for designing the metasurface phase, and use it to construct a proof-of-principle beam shaping metasurface. We demonstrate the optical performance of the fabricated sample, and perform numerical experiments to investigate the quality of the metasurface design. Our approach bridges the fields of non-imaging optics and metaoptics, and may enable metasurface applications for shaping coherent and incoherent light.
\newpage

\section*{Introduction}
Artificial light plays an important role in daily life, with applications ranging from illumination to communication. Over the past decade, light-emitting diodes (LEDs) have become the dominant technology driving advancements in illumination due to their low power consumption, small footprints and high efficacy~\cite{2Lane2023Lighting,3Nair2021FundamentalsLEDs}. LEDs typically emit light in a broad angular distribution, such as a Lambertian distribution which is equally bright in all directions. It is thus often necessary to use optical elements to shape the intensity distribution of LEDs for different applications. There are many different strategies to this end, including Fresnel and total internal reflection lenses~\cite{16Teupner2015OptimizationLights}, polymer coatings~\cite{18Lin2019ShapingLattices}, lens arrays~\cite{17Zhang2017AdjustableArray}, reflectors~\cite{20Ying2021SingleLight} and multilayer thin films~\cite{19Wankerl2022DirectionalFilms}. However, the most versatile approach to design optics for LEDs is found in the field of freeform lenses, which offers design flexibility and can be customized for specific applications~\cite{21Wu2018DesignOptics}. In illumination applications, freeform lenses are designed to be non-imaging, and the design problem is formulated as an inverse problem. By knowing the intensity distribution incident on the lens and the desired output intensity at a certain plane, it is possible to calculate the lens shape that will perform the correct mapping. There are two main strategies for solving the problem~\cite{22Brix2015DesigningEquations}, namely iterative ray tracing schemes~\cite{23Ding2008FreeformIllumination,24Sun2009Free-formApplications,25Fournier2010FastMaps,26Chen2012FreeformIllumination} and the non-linear differential equation approach~\cite{27Ries2002TailoredSurfaces,28Wu2013FreeformEquation,29Gutierrez2018ReflectionEquations}. The latter does not require a priori knowledge of the lens shape and has been proven to work well for a wide range of light sources and target patterns~\cite{30Wu2014InfluenceDesign,31Mao2014Two-stepSource}. However, freeform lenses suffer from a few key aspects: they are bulky like all traditional optical elements and, depending on the application, it might be difficult to reach the necessary precision in the fabrication~\cite{34Wang2009EffectLenses,35Fang2013ManufacturingOptics,36Nie2021RBFGeneration,37Chen2024FabricationRuling}.

A flat alternative to traditional optics is emerging in the field of metaoptics. Optical metasurfaces can manipulate the properties of light through carefully engineered nanostructures, and thus achieve similar functionality as traditional optics. This field is well established for coherent light sources~\cite{38Paniagua-Dominguez2014AAperture,39Zhou2017EfficientLight,40vandeGroep2020ExcitonLens,41Lawrence2020HighMetasurfaces,42Cai2021Dual-FunctionalMetasurfaces}, but is still under development for incoherent sources~\cite{43So2023RevisitingBeyond,44Khaidarov2020ControlMetasurfaces,45Bayati2022InverseVisible,46Mukherjee2023PartiallyMeta-Optics}. A big advantage of metaoptics, aside from the small footprint, is that once a reliable and high-quality design platform for the nanostructures is developed, one can implement complex phase profiles without additional fabrication challenges. 

Here we present a way to bridge the fields of non-imaging freeform lenses and metaoptics, in order to create a design platform for non-imaging metasurfaces that works with incoherent sources such as LEDs. We show the derivation of a theoretical framework for calculating metasurface phase profiles, both in 1D and 2D, based on the optimal transport formulation from non-imaging optics. We present \texttt{MetaShape}, a numerical implementation of our framework, and combine it with the open-source beam propagator \texttt{diffractsim} to simulate the output of our designed metasurfaces. Finally, we design, fabricate, and characterize a proof-of-principle metasurface that shapes the intensity from a collimated laser diode into a ring pattern.

\section*{Theoretical framework}
\begin{figure}[t!]
    \centering
    \includegraphics[width=\textwidth]{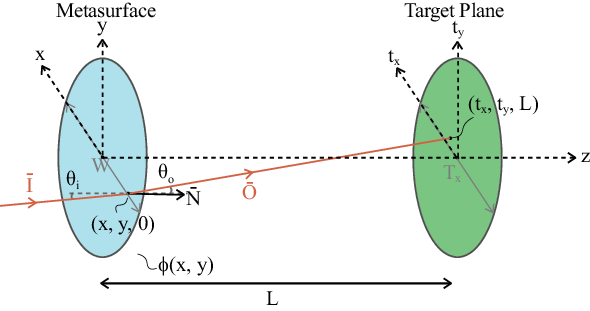}
    \caption{Illustration of the two-dimensional problem. The ray $\Tilde{I}$ is incident at an angle $\theta_i$ on a metasurface in the point $P = (x,y,z=0)$. The refracted ray $\Tilde{O}$ propagates at an angle $\theta_o$ to the target plane, where it is incident on the point $P_\textrm{T} = (t_x,t_y,z=L)$. The incident and target intensity profiles are known, and we aim to calculate the metasurface phase profile $\phi(x,y)$.}
    \label{fig:Schematic}
\end{figure}

Based on the formulation of optimal transport from non-imaging optics, we derive a theoretical framework for calculating the phase profile of a metasurface. In previous work, we have shown how to calculate the one-dimensional phase profile of a metasurface that shapes a collimated beam~\cite{Nielsen2023Non-imagingShaping}, and how to expand the model to calculate two-dimensional phase profiles in cylindrically symmetric cases \cite{Nielsen2024Cylindrically-symmetricMetasurfaces}. Here we combine these results and expand the model to account for non-collimated beams as well. Let us first consider a beam of light traveling in the $\hat{z}$ direction with a one-dimensional intensity profile $I(x)$. The beam illuminates a metasurface of width $W$, which is positioned at $z = 0$ (see Fig.~\ref{fig:Schematic}). The metasurface maps the incident beam to a one-dimensional target intensity profile $E(t_x)$ of width $T_x$ on a target plane positioned at $z = L$. We use the coordinates $x$ and $t_x$ on the metasurface and target planes, respectively. First we require the mapping of the intensity profiles to conserve energy, i.e., that all power incident on the metasurface is mapped to the target pattern. In the one-dimensional case this means
\begin{align}\label{eq:1D_power}
    \int_W I(x) \text dx = \int_{T_x} E(t_x) \text dt_x = \int_W E(t_x)\left|\frac{\partial t_x}{\partial x}\right| \text dx,
\end{align}
where $\left|\frac{\partial t_x}{\partial x}\right|$ is the determinant of the Jacobian matrix describing the mapping. In the two-dimensional cylindrically symmetric case, we need to rewrite the integral in terms of cylindrical coordinates, leading to 
\begin{align}\label{eq:2d_power}
    \int_W I(x) |x| \text dx = \int_{T_x} E(t_x) |t_x| \text dt_x = \int_W E(t_x)|t_x|\left|\frac{\partial t_x}{\partial x}\right| \text dx,
\end{align}
where $|x|$ and $|t_x|$ represent the radial distance from the center to an arbitrary point on the metasurface and target planes, respectively~\cite{Nielsen2024Cylindrically-symmetricMetasurfaces}. Assuming local power conservation, Eqs.~(\ref{eq:1D_power}-\ref{eq:2d_power}) result in the differential equations
\begin{align}\label{Eq:1D_diff}
    \text{1D: } \frac{\partial t_x}{\partial x} &= \frac{I(x)}{E(t_x)}, \\
    \text{2D: } \frac{\partial t_x}{\partial x} &= \frac{I(x)|x|}{E(t_x)|t_x|}\label{Eq:2d_diff}.
\end{align}
Here, we have lifted the absolute value from the Jacobian and chosen the positive $(+)$ sign in the equation. This choice comes with the boundary condition $t_x\left(x = \pm\frac{W}{2}\right) = \pm \frac{T_x}{2}$ as discussed in our previous work \cite{Nielsen2023Non-imagingShaping}.

Next, we consider the refraction performed by the metasurface on the incident beam. A refracted ray is defined by its wave vector $\textbf{k}_o = (k_{o,x},k_{o,z})$, and the unit vector of the refracted ray is defined as $\textbf{O} = \frac{\textbf{k}_o}{|k_o|} = (O_x,O_z)$. From geometrical considerations, we find the relationship between the metasurface coordinate $x$ and target plane coordinate $t_x$ to be
\begin{align}\label{eq:CoordinateRelationship}
    t_x = x + L\frac{O_x}{O_z}.
\end{align}
The refraction caused by the metasurface can be described with the generalized law of refraction which for non-normal incidence can be written as \cite{Moreno2020NonimagingMetaoptics,Aieta2012Out-of-PlaneDiscontinuities}
\begin{align}
    \frac{\partial \Phi}{\partial x} = k_{o,x}-k_{i,x} = k_0n_o\sin\theta_o - k_0n_i\sin\theta_i.
\end{align}
Here, $\frac{\partial \Phi}{\partial x}$ is the phase gradient of the metasurface, $n_i$ and $n_o$ are the refractive indices of the media before and after the metasurface, respectively, $\theta_i$ is the incidence angle of the ray and $\theta_o$ is the corresponding angle of the refracted ray. This leads to an expression for the unit vector of the refracted ray
\begin{align}
    O_x &= \frac{k_{o,x}}{n_ok_0} = n_i\sin\theta_i+\Tilde{\Phi}_x, \\
    O_z &= \frac{k_{o,z}}{n_ok_0} = \frac{\sqrt{(n_ok_0)^2-k_{o,x}^2}}{n_ok_0} = \sqrt{1-(n_i\sin\theta_i+\Tilde{\Phi}_x)^2},
\end{align}
where we use the notation $\Tilde{\Phi}_x = \frac{1}{n_ok_0}\frac{\partial \Phi}{\partial x} $. Using these results in Eq.~(\ref{eq:CoordinateRelationship}), we can derive an expression for the wavelength-independent phase gradient of the metasurface
\begin{align}\label{Eq:gradient}
    \Tilde{\Phi}_x = \frac{t_x-x}{\sqrt{L^2+(t_x+x)^2}}-n_i\sin\theta_i.
\end{align}
where $t_x(x)$ is obtained by solving Eq.~(\ref{Eq:1D_diff}) in 1D, or Eq.~(\ref{Eq:2d_diff}) in 2D. Note that for a normally incident beam the last term vanishes and we retrieve the result presented in Ref. \cite{Nielsen2023Non-imagingShaping}. To calculate the wavelength-dependent phase profile, we need to multiply the result with the specific wave number $k_0=2\pi/\lambda$ and integrate the phase gradient. This result is valid in both the one- and two-dimensional cases. To obtain the two-dimensional phase profile in a cylindrically symmetric case, we rotate the 1D phase gradient to get the full 2D profile. Once the phase profile is calculated, we use the angular spectrum method, implemented in the Python toolbox \texttt{diffractsim}, to propagate the electric field and calculate the resulting intensity pattern on the target plane. In order to mimic a real metasurface with a certain pixel size, we implement the phase profile on a grid with a spacing corresponding to the metasurface period, and set the intensity and phase to zero outside of the metasurface area, i.e., effectively putting an aperture around the metasurface.
As a final note, this derivation requires that the intensity distributions be given in terms of the metasurface and target plane coordinates $x$ and $t_x$, respectively. For incident beams that are described by an angular intensity distribution, it is useful to project the distribution onto the metasurface plane to keep the equations unchanged. To this end, we use a stereographic projection, which has been successfully used in other non-imaging designs \cite{Feng2016FreeformMap}. We describe this procedure in more detail later.

In some cases, it is possible to solve Eqs.~(\ref{Eq:1D_diff}-\ref{Eq:2d_diff}) analytically, as previously demonstrated for a normally-incident beam in one dimension~\cite{Nielsen2023Non-imagingShaping}. However, most cases require a numerical solution. We therefore introduce \texttt{MetaShape}, a numerical implementation of our framework that serves as an easy-to-use design tool for non-imaging metasurfaces. \texttt{MetaShape} is implemented in Python and is available on Github~\cite{MetaShape}. The differential equations presented in Eqs.~(\ref{Eq:1D_diff}-\ref{Eq:2d_diff}) are boundary value problems. However, since we have limited ourselves to symmetric problems, we can solve them as initial value problems using one of the boundary points as the initial condition and observe that the second boundary condition is satisfied. We then use the result to solve Eq.~(\ref{Eq:gradient}) and use numerical integration to obtain the phase profile. In the 2D case, we interpolate the 1D phase profile to calculate the full phase profile on a 2D grid. In order to test the results of our phase design, we have built \texttt{MetaShape} to work in combination with the beam propagation package \texttt{diffractsim}~\cite{Herrezuelo2022Diffractsim:Simulator}. In the following, we use \texttt{MetaShape} to design and simulate actual working samples.

\begin{figure}[t!]
    \centering
    \includegraphics[width=\textwidth]{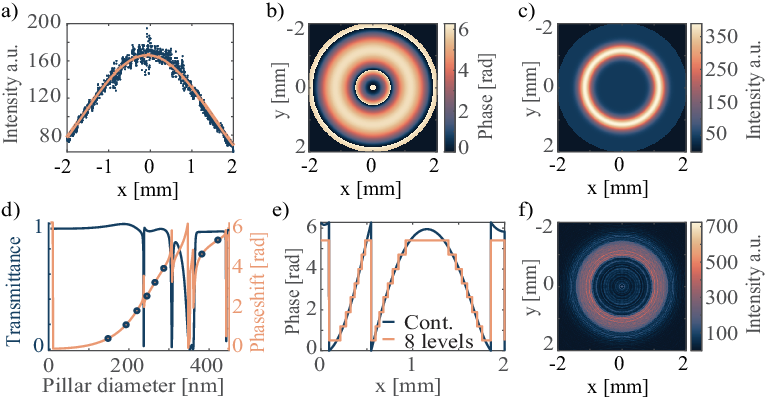}
    \caption{Design and simulation of the non-imaging metasurface. (a) The incident laser intensity profile is fitted using a Gaussian function. (b) Calculated metasurface phase profile using \texttt{MetaShape} which shapes the incident Gaussian beam into the ring pattern shown in (c). (d) Look-up table for silicon nitride nanopillars calculated using full-wave simulations. (e) The calculated phase profile is discretized into eight phase levels based on the look-up table. (f) Simulated target pattern using the angular spectrum method accounting for the finite size and phase discretization of the metasurface.}
    \label{fig:Experiment_part1}
\end{figure}

\section*{Experimental verification}
We verify our theoretical framework and numerical implementation by fabricating and characterizing a proof-of-principle metasurface. Our illumination source is a diode laser with a wavelength of $\lambda = 635$~nm (Thorlabs PL202). To obtain a Gaussian intensity profile at the metasurface plane, we clean up the laser mode using an iris at the laser exit and by propagating the beam for 3.3~m. A cross section of the beam spot along with a Gaussian fit are shown in Fig.~\ref{fig:Experiment_part1}a. The Gaussian fit in this case is $I(x) = 166.1\cdot\textrm{exp}\left(-\left(\frac{x\textrm{[mm]}}{2.205\textrm{
mm}}\right)^2\right)$. We want to map the intensity of the incident beam to a ring pattern (Fig.~\ref{fig:Experiment_part1}c), which in 1D is a double Gaussian pattern, at a distance of $L = 20$ cm from the metasurface. Using \texttt{MetaShape}, we calculate the 2D phase profile which performs the desired mapping (Fig.~\ref{fig:Experiment_part1}b). The metasurface and target patterns are both 4~mm in diameter.

In order to translate the calculated phase profile into a metasurface design, we calculate a look-up table for the meta-atoms that will make up the metasurface using COMSOL Multiphysics (Fig.~\ref{fig:Experiment_part1}d). We realize the meta-atoms using silicon nitride nanopillars with a fixed height of 697~nm~(Ref.~\cite{Zhan2016Low-ContrastOptics}). The pillars are placed in an array with a period of 500~nm. The diameter of the pillars determines the local phase shift, and we cover most of the $2\pi$ phase range by varying the diameter from 0 to 450~nm. By choosing eight evenly spaced phase levels, we can cover the phase range $0.51$ rad to $5.41$~rad, while maintaining a transmission close to unity (Fig.~\ref{fig:Experiment_part1}d). This results in pillar diameters from 148~nm to 422~nm. Ideally we would use more phase levels to better sample the phase, however, due to grating effects in the arrays and limitations from the fabrication we choose to use only eight levels. Figure~\ref{fig:Experiment_part1}e shows a comparison between the calculated continuous phase profile and the discretized phase profile using eight phase levels. We use \texttt{diffractsim} with the eight-level phase profile to simulate the beam shaping performed by our metasurface, with the resulting target pattern shown in Fig.~\ref{fig:Experiment_part1}f. While the simulated pattern overall matches with the design pattern (Fig.~\ref{fig:Experiment_part1}c), it is clear that there are notable deviations. These deviations are due to diffraction effects and the discretization of the phase profile, both of which are not accounted for in our theoretical framework. We discuss this in more detail in the next section.

\begin{figure}[t!]
    \centering
    \includegraphics[width=\textwidth]{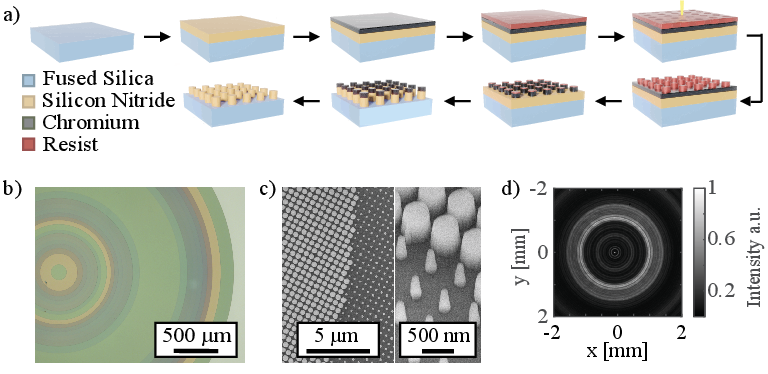}
    \caption{Fabrication of non-imaging metasurface. (a) Illustration of the cleanroom process flow. (b-c) The fabricated metasurface is characterized with optical microscopy and scanning electron microscopy, respectively. (d) Measured intensity pattern at the target plane ($z=L=20$~cm) due to the beam-shaping metasurface. The measurement is in good agreement with the simulated pattern (shown in Fig.~\ref{fig:Experiment_part1}f).}
    \label{fig:Experiment_part2}
\end{figure}

We fabricate the metasurface in a cleanroom facility using the process flow illustrated in Fig.~\ref{fig:Experiment_part2}a. Our substrate is a 4-inch fused silica wafer with a thickness of 500~$\mu$m. On top of the wafer we deposit 697~nm silicon nitride using PECVD with a mixed frequency recipe in a SPTS Multiplex PECVD system, which yields a refractive index of 2.07 at a wavelength of $\lambda = 635$~nm \cite{Beliaev2022OpticalDeposition}. The thickness and refractive index of the film is measured with ellipsometry. The wafer is then cut into $2\times2$~cm$^2$ chips for further processing. Next, we deposit 80~nm of chromium on top of the chip using electron beam evaporation at a rate of 1~Å/s in a Temescal FC-2000 e-beam evaporator from Ferrotec. On top of the chromium we spin coat 200~nm of the negative resist AR-N 7520.11.new from All Resist, and expose our metasurface design in the resist using a JEOL JBX-9500FS electron-beam lithography system at a current of 0.8~nA and an acceleration voltage of 100~kV. The pattern is developed for 90~s in AZ MIF 726 developer (TMAH) and subsequently dipped in water. The pattern is transferred into the chromium layer with an anisotropic etch in a Pro ICP from STS at room temperature using a mixture of chlorine and oxygen gases with 23\% oxygen. We then use chromium as an etch mask for silicon nitride, which we etch in a STS MESC Multiplex ICP Advanced Silicon Etcher at room temperature with a mixture of trifluoromethane (CHF$_3$) and hydrogen gases in a 1:1 ratio. Any remaining chromium is removed by dipping the chip in Chrome Etch 18 from OSC GmbH. An optical micrograph and scanning electron microscopy image of the final sample are shown in Fig.~\ref{fig:Experiment_part2}b-c, respectively.

Finally, we characterize the performance of the metasurface by placing it in the beam path of our laser module. The target pattern produced by the metasurface at a distance of 20~cm is recorded with a DMK 33UX178 gray scale camera from The Imaging Source, and the result is shown in Fig.~\ref{fig:Experiment_part2}d. The performance of the metasurface is in good agreement with our simulation (Fig.~\ref{fig:Experiment_part1}d), providing an experimental verification of our theoretical framework and \texttt{diffractsim} as a beam propagator tool. 

\begin{figure}[t!]
    \centering
    \includegraphics[width=\textwidth]{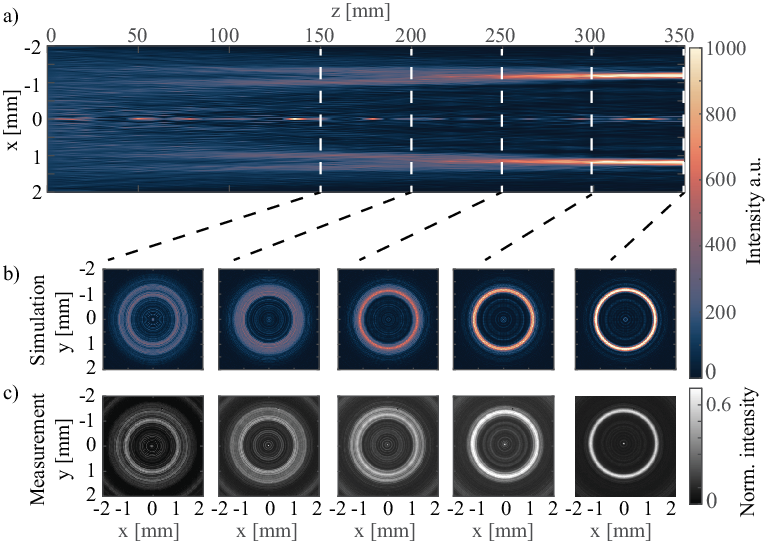}
    \caption{Beam propagation after interacting with the metasurface. (a) Beam intensity along the propagation direction from $z = 0$~cm to $z = 35$~cm. (b-c) Simulated and measured target patterns at distances $z = 15, 20, 25, 30, 35$ cm from the metasurface plane, respectively. The beam is focused after the target plane and acquires an increasingly Gaussian shape. The width of the target ring pattern is best maintained at the designed distance $z = L = 20$~cm.}
    \label{fig:PropagationDistance}
\end{figure}

Our sample is designed to produce a ring-shaped pattern at a distance of $z = 20$~cm (see Fig.~\ref{fig:Experiment_part1}c) by bending the incident light rays and effectively focusing the rays to form the ring. Figure~\ref{fig:PropagationDistance}a shows the intensity of the beam as it propagates from $z = 0$ cm to $z = 35$~cm, and clearly shows the focusing effect. To better visualize the beam shaping, we compare the simulated (Fig.~\ref{fig:PropagationDistance}b) and the measured (Fig.~\ref{fig:PropagationDistance}c) intensity patterns at the distances $z = 15, \ 20, \ 25, \ 30, \ 35$ cm, respectively. We see that the intended intensity map at $z = 20$ cm is not placed in the apparent focal point of the beam. As we move further away, the ring pattern becomes more focused and takes an increasingly Gaussian-like shape, although the width of the ring becomes much more narrow than in the original design. The width of the ring is closer to the intended design at the designed distance $z = 20$~cm, meaning that our mapping preserves the width of the design, but additional propagation is needed to achieve the Gaussian shape. It is worth noting that our choice of boundary conditions ensures that we produce the desired pattern before the focal point, as discussed in our previous paper Ref.~\cite{Nielsen2023Non-imagingShaping}. This also suggests that in the example of Figure~\ref{fig:PropagationDistance}, we might recover the desired intensity profile at longer distances on the opposite side of the focal point of the beam. In contrast, a different choice of boundary conditions would effectively move the focal point of the beam closer to the metasurface, yet we would still recover our designed pattern at target plane $z = 20$~cm. 

\section*{Discussion}
To gain a better understanding of the underlying reasons that make our target intensity (see Fig.~\ref{fig:Experiment_part1}c) differ from the simulated (and measured) intensity profile (see Fig.~\ref{fig:Experiment_part1}f and Fig.~\ref{fig:Experiment_part2}d, respectively), we carry out a series of numerical experiments using \texttt{diffractsim}. To this end, we investigate the mapping of a light source with a Lambertian intensity distribution $I(\theta) = I_0\cos(\theta)$ to a constant intensity at the target plane. This mapping is chosen to provide a clear demonstration of the effects of having non-zero intensity at the edge of the metasurface and makes it easy to spot deviations from the intended constant target pattern. In addition, the mapping demonstrates the capabilities of \texttt{MetaShape} to handle an incident beam with an angular distribution, which is relevant for LED sources.

\begin{figure}[t!]
    \centering
    \includegraphics[width=\textwidth]{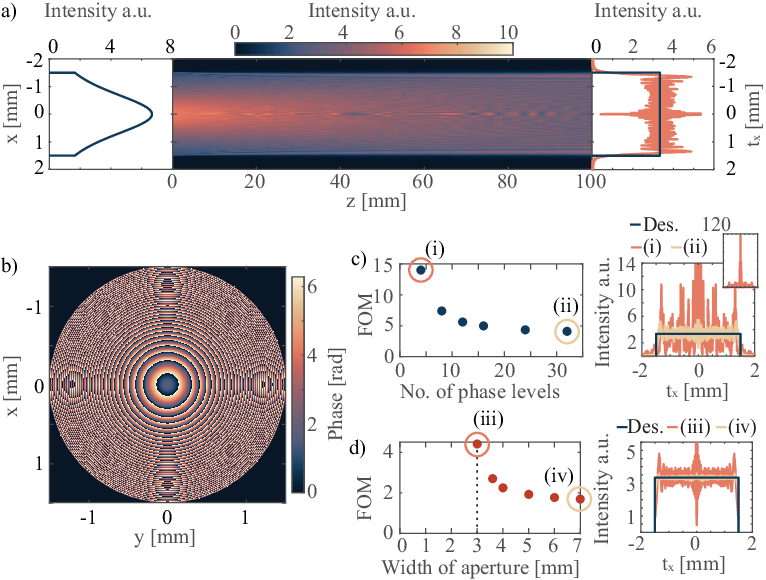}
    \caption{Phase design and beam propagation of a Lambertian source to a constant intensity target pattern. (a) Cross sections of the incident intensity profile and the designed and simulated target intensity profiles, as well as the beam intensity along the propagation path. (b) The calculated phase profile wrapped in the interval $[0,2\pi]$. (c) Impact of the number of discretized phase levels on the quality of the target pattern, computed with the FOM defined in Eq.~(\ref{eq:FOM}). In the left panel, FOM is given as a function of the number of phase levels. In the right panel, the intensity cross sections of the design target, as well as points (i) and (ii) are given. Points (i) and (ii) refer to the phase profiles with 4 and 32 phase levels, respectively. d) Impact of the aperture width on the target pattern, quantified via the same FOM. In the left panel, FOM is given as a function of the width aperture. In the right panel, the intensity cross sections of the design target, as well as points (iii) and (iv) are given. Points (iii) and (iv) refer to the calculations made with apertures of 3 and 7~mm, respectively.} 
    \label{fig:Theory example}
\end{figure}

Since our theoretical model uses as input the intensity distribution at the metasurface plane, i.e., $I(x)$, we first project the angularly dependent intensity distribution of a Lambertian source into an intensity profile at the metasurface plane. We achieve this by applying a stereographic projection~\cite{Feng2016FreeformMap}, where we consider a unit sphere with its center located at $(x,y,z) = (0,0,0)$, and let $(u,v)$ denote the Cartesian coordinates describing the plane cutting through the center of the sphere at $z = 0$. Using the south pole convention, we can write the flux, $\textrm{d}F$, through the line on the plane for which $v = 0$ and the corresponding projected intensity on the $(u,v = 0)$ line, $E(u)$, as
\begin{align}
    \textrm{d}F = I(u)\left(\frac{2}{1+u^2}\right)^2 \textrm{d}u \Rightarrow E(u) = I(u)\left(\frac{2}{1+u^2}\right)^2,
\end{align}
with $I(u)$ being the incident intensity distribution in terms of the coordinate $u$. Since $(u,v)$ exist within the unit sphere, we have the relation $u = x/\frac{W}{2}$ between the coordinate $u$ and the metasurface coordinate $x$. We can now do a substitution of variables with $\textrm{d}u = \frac{2}{W}\textrm{d}x$, and by simultaneously substituting $\cos(\theta) = \frac{L_s}{\sqrt{L_s^2+x^2}}$, we arrive at the projected intensity on the metasurface due to a Lambertian angular distribution
\begin{align}
    E(x) = I_0\frac{L_s}{\sqrt{L_s^2+x^2}}\left(\frac{2}{1+(2x/W)^2}\right)^2\frac{2}{W},
\end{align}
where $L_s$ is the absolute distance from the light source to the metasurface. For the mapping shown in Fig.~\ref{fig:Theory example} we use $L_s = 5$~cm. We assume that the light source, the metasurface and the target plane are centered with respect to each other. The metasurface in this case is 3~mm wide, and we aim to achieve a constant intensity at $L = 10$ cm. The mapping is summarized in Fig.~\ref{fig:Theory example}, where the designed phase profile (Fig.~\ref{fig:Theory example}b) has been obtained using our framework given by Eqs.~(\ref{Eq:2d_diff}) and (\ref{Eq:gradient}). The phase profile is discretized every 500~nm with phase values spanning the range $\left[0, 2\pi \right]$.

A cross section of the beam intensity along the propagation direction is shown in Fig.~\ref{fig:Theory example}a, as well as the incident, target and simulated transverse intensities at $z=0$ and $z=100$ mm, respectively. We observe that the simulated target pattern is noisy, suggesting that the phase design is not perfect. We suspect diffractive effects to be the main cause of this difference, and next we investigate two parameters which might improve the beam shaping: i) The number of phase levels used to discretize the phase profile and ii) the ratio between the beam size and the width of the aperture in the simulation. To quantify the quality of the mapping, we use the following figure-of-merit (FOM)
\begin{align}
    \textrm{FOM} = \int_{T_x} \left|E_\textrm{sim}-E_\textrm{design}\right| \text d t_x,
    \label{eq:FOM}
\end{align}
where $E_\textrm{sim}$ and $E_\textrm{design}$ are the simulated and designed intensity target patterns, respectively. Figures~\ref{fig:Theory example}c-d show the quality of the mapping as a function of number of phase levels and width of the aperture, respectively, as well as a cross section of the target pattern for the highest and smallest FOM. Note that for the design of the fabricated metasurface we used eight phase levels and an aperture width corresponding to the width of the metasurface ($W = 4$~mm), which effectively cuts the incident Gaussian beam at a transverse position which is smaller than the beam waist. In both cases, we assume a metasurface period, and thus grid spacing, of 500 nm. Keeping the aperture at the same width of the metasurface (and therefore cutting more than 32\% of the beam power), we see that a larger number of phase levels will yield better results to a certain point, as might be expected. However, an increasing amount of phase levels cannot completely remove the noise. Using the finest possible resolution for the phase, we now increase the size of the aperture and see a clear improvement of the simulated target pattern, suggesting that diffraction at the edges of the aperture greatly affects the beam shaping. In previous work, we have seen that incident intensities that approach zero at the edge of the metasurface yield target patterns with very low noise \cite{Nielsen2023Non-imagingShaping}, which further supports our finding. We also observe a small intensity spike in the center of the pattern in both the simulated intensity plots (Fig.~\ref{fig:PropagationDistance} and Fig.~\ref{fig:Theory example}a,c-d) and the experimental intensity measurements (Fig.~\ref{fig:Experiment_part2}d). This occurs due to diffraction from the finite size of the metasurface, since the intensity peak diminishes as we widen the aperture used in the simulations (Fig.~\ref{fig:Theory example}d). Thus, we conclude that our metasurface design strategy requires an incident beam whose intensity distribution is mostly incident on the metasurface to minimize diffractive effects due to the finite size of the metasurface. In addition, maximizing the number of phase levels is beneficial to reduce the noise level and achieve the designed intensity pattern. 

\section*{Conclusion}
In conclusion, we have presented a theoretical framework as well as a numerical implementation (\texttt{MetaShape}) that allows us to design non-imaging metasurfaces for intensity shaping of two-dimensional cylindrically symmetric beams. We demonstrate the capabilities of the \texttt{MetaShape} Python tool for both design and numerical experiments, and present a proof-of principle design and implementation of a phase profile in a physical sample. Our measurements show good agreement with our simulations, but both deviate from the original design. We investigate the cause of the noise in the produced target patterns, and argue that it originates from diffraction effects. Our results provide a bridge between the fields of non-imaging freeform optics and metaoptics and present a new versatile design method for non-imaging metasurfaces, with potential application in lighting.

\subsection*{Funding}
S.~R. and X.~Z.-P. acknowledge funding from VILLUM FONDEN (VIL50376).

\subsection*{Disclosures}
The authors declare no conflicts of interest.

\subsection*{Data availability} 
Data underlying the results presented in this paper are not publicly available at this time but may be obtained from the authors upon reasonable request.


\bibliography{References}
\newpage

\end{document}